\documentclass[]{spie}  

\usepackage{amsmath,amsfonts,amssymb}
\usepackage{graphicx}
\usepackage[colorlinks=true, allcolors=blue]{hyperref}

\title{Air, Telescope, and Instrument Temperature Effects on the Gemini Planet Imager's Image Quality }

\author[a]{Melisa Tallis}
\author[b]{Vanessa P. Bailey}
\author[a]{Bruce Macintosh}
\author[a]{Jeffrey K. Chilcote}
\author[c]{Lisa A. Poyneer}
\author[a]{Jean-Baptiste Ruffio}
\author[d]{Thomas L. Hayward}
\author[e]{Dmitry Savransky}
\author[ ]{the GPI Team}

\affil[a]{Kavli Institute for Particle Astrophysics \& Cosmology, Physics Department, Stanford University, Stanford, CA, 94305, USA}
\affil[b]{Jet Propulsion Laboratory, California Institute of Technology, Pasadena, CA, 91109, USA}
\affil[c]{Lawrence Livermore National Laboratory, 7000 East Ave, Livermore, CA, 94550, USA}
\affil[d]{Gemini Observatory, Casilla 603, La Serena, Chile}
\affil[e]{Sibley School of Mechanical and Aerospace Engineering, Cornell University, Ithaca, NY, 14853, USA}

\authorinfo{Further author information:
\\Melisa Tallis: E-mail: melisatallis@gmail.com
\\Bruce Macintosh: E-mail: macint@stanford.edu}

\begin{document} 
\maketitle
\begin{abstract}
The Gemini Planet Imager (GPI) is a near-infrared instrument that uses Adaptive Optics (AO), a coronagraph, and advanced data processing techniques to achieve very high contrast images of exoplanets. The GPI Exoplanet Survey (GPIES) is a 600 stars campaign aiming at detecting and characterizing young, massive and self-luminous exoplanets at large orbital distances ($>5~au$). Science observations are taken simultaneously with environmental data revealing information about the turbulence in the telescope environment as well as limitations of GPI's AO system. Previous work has shown that the timescale of the turbulence, $\tau_0$, is a strong predictor of AO performance, however an analysis of the dome turbulence on AO performance has not been done before. Here, we study correlations between image contrast and residual wavefront error (WFE) with temperature measurements from multiple locations inside and outside the dome. Our analysis revealed GPI's performance is most correlated with the temperature difference between the primary mirror of the telescope and the outside air. We also assess the impact of the current temperature control and ventilation strategy at Gemini South (GS). 
\end{abstract}


\section{INTRODUCTION}

Direct imaging of exoplanets is a technique to spatially resolve exoplanets' light from that of their host star. It provides a unique window to detect and characterize exoplanets at planet semi-major axis difficult to access with indirect methods ($>5~au$).
Each newly imaged planet helps to constrain planet populations and provides a laboratory to study the formation, evolution, and physics of young Jupiters\cite{Bowler2016ImagingPlanets}. The Gemini Planet Imager Exoplanet Survey (GPIES)\cite{Macintosh2008TheConstruction,MacintoshPROCEEDINGSCommissioning} is a campaign started in January 2014 that aims to image 600 young, nearby stars with the GPI instrument. 
\newline
\indent 
Direct imaging is challenging; indeed, a young Jupiter is about a million times fainter than its host star at near-infrared wavelengths and its projected separation is expected to be less than 0.5''. In theory, a telescope equipped with a large enough mirror would have high enough angular resolution to image exoplanets. In practice, turbulence in Earth’s atmosphere limits imaging performance, even under good observing conditions. Direct imaging instruments like the Gemini Planet Imager (GPI) take all of these factors into consideration. 
\newline
\indent 
AO systems were developed to improve imaging performance of ground based telescopes, and are now being built into the world's largest telescope facilities. These systems continuously measure and correct for atmospheric effects by deforming the shapes of mirrors in the optical path. For a complete description of GPI AO hardware and software, the reader may refer to Poyneer\cite{PoyneerPerformanceSystem} (2016). Unfortunately, AO systems are limited by many factors including a time-lag between a measurement of the atmosphere and the associated correction. Inaccurate corrections result in noisy artifacts (speckles) that can overwhelm the signal from a faint planet. 
\newline
\indent 
The large, uniform GPIES dataset enables the study of the relationship between various environment and telescope parameters and GPI performance. Previous work showed that the turbulence coherence timescale ($\tau_0$) is the strongest predictor of GPI imaging performance\cite{PoyneerPerformanceSystem,Bailey2016StatusSystem}. However, there were several instances when the atmospheric seeing at Cerro Pachon was very good, but the quality of our images remained mediocre. We hypothesize that the image deterioration was induced by turbulence in the telescope environment (local seeing). In this proceedings we present results from our study of local seeing at the Gemini South 8-m telescope (GS) at Cerro Pachon. The goal of this study is to identify the sources of local seeing and to quantify their effects. We assess the current temperature control and ventilation strategy at GS and recommend improvements so that GPI may take full advantage of the best atmospheric seeing conditions at Cerro Pachon.
\label{sec:intro}  

\section{Local Seeing}
Turbulence can come from various sources, including the free atmosphere, turbulent air flows generated by the telescope dome, and temperature gradients in the instrument's immediate surroundings. We consider that the total amount of image spread (seeing) due to the atmosphere and dome\cite{Racine1991MirrorCFHT} is:
\begin{equation}
\label{eq:fov}
S^\frac{5}{3} = S_A^\frac{5}{3} + S_L^\frac{5}{3}
\end{equation}
where $S$ is the total seeing, $S_A$ is the seeing inherent to the atmosphere, and $S_L$ is the local seeing associated with the dome. When the site seeing is bad, there is little that can be done to improve matters. But when seeing is very good, then $S_L$ may dominate. Studies aimed at identifying and quantifying local seeing effects have been done by several groups, with the goal to improve the design and management of observatories in order to optimize the telescope's performance. For example, one of the first local seeing studies was done by Hoag\cite{HoagA.A.PriserJ.B.InstallationReflector} (1967) during the commissioning of the Flagstaff telescope. He observed the amount of image spread due to temperature gradients using temperature sensors and photographic plates. Murdin and Bingham\cite{MurdinP.andBinghamR.G.AnSiz} (1975) measured the total amount of heat excess allowed in the Royal Greenwhich observatory before image degradation became significant. Gillingham\cite{Gillingham1983PROCEEDINGSTelescope} (1978) investigated  the frequency of temperature fluctuations at the Anglo-Australian Telescope. Some of the most recent observations were made by Racine\cite{Racine1991MirrorCFHT} (1991) at the Canada France Hawaii Telescope. The amount of image spread generated from temperature gradients near the instrument is proportional to the square of the temperature difference between two points\cite{Fried1966LimitingAtmosphere}. Thus mirror seeing expressed in units of residual wavefront error (WFE) would be  proportional to:
\begin{equation}
\label{eq:fov}
\sigma_{WFE}^2 \propto \Delta{T}_M^2 
\end{equation}
where $\sigma_{WFE}$ is the WFE and $\Delta{T}$ is the temperature difference between the primary mirror and the ambient air.   

\section{The Data}
Since the start of the GPIES campaign in 2014, GPI has accumulated over 25,000 science images and over 2,000 telemetry sets. Science images are spectral cubes taken with 60 second image exposure. A GPIES campaign observation consists of approximately 20-40 single frames per target. Image quality metrics and telemetry information are matched with each image in the GPIES database; this database allows for the investigation of trends in the instrument's AO performance. 
\newline
\indent An important measure of the instrument's performance is ``contrast,'' the flux ratio between a the faintest detectable planet and its host star, at a given projected separation from the host star. A higher contrast corresponds to detecting fainter planets. GPI pipeline\cite{Perrin2014GeminiPipeline} estimates contrast by computing the standard deviation of the intensity in an annulus and multiplying it by five. This defines the detection threshold for the faintest planet. In this analysis we study two types of contrast. ``Raw contrast'' refers to a single-frame contrast, whereas ``final contrast'' refers to the combined and post processed contrast of a full observing sequence. Final contrast is typically 10-100 times more sensitive than raw contrast. 
\newline
\indent
GPI's contrast performance is directly linked to its AO performance. A direct measurement of the AO system's performance is the residual wavefront error $(\sigma_{WFE})$. Defined as the total error measured by the wavefront sensor after applying a correction with the deformable mirror. An instantaneous approximation of the total residual WFE, based on the RMS subaperture slopes, is saved to the header of each frame. We recalibrate the header WFE value using a linear fit derived from a subset of images that had contemporaneous residual phase maps reconstructed from full AO telemetry. 

\begin{figure} [ht]
   \begin{center}
   \begin{tabular}{c} 
   \includegraphics[height=7cm]{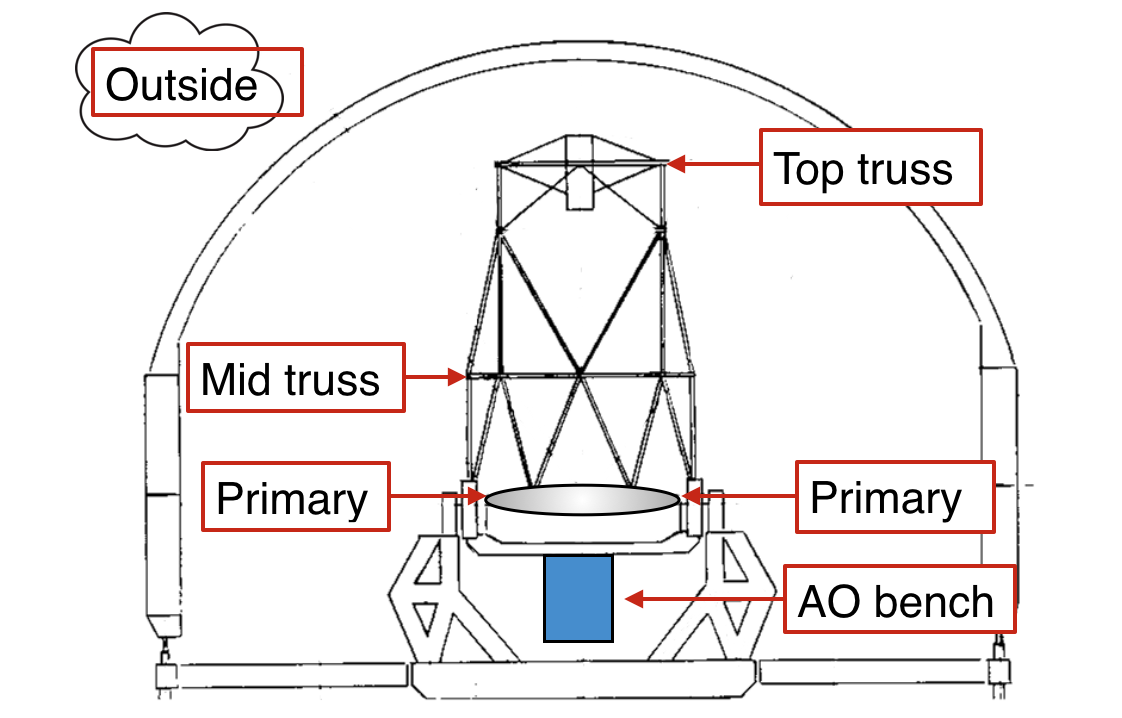}
   \end{tabular}
   \end{center}
   \caption[example] 

   { \label{fig:schematic} 
Schematic of GS. Arrows mark the location of temperature sensors.}
   \end{figure}

The telemetry data are recorded by the observatory's Differential Image Motion Monitor (DIMM) and Multi-Aperture Scintillation Sensor (MASS). We use these measurements to select datasets when atmospheric seeing is good.   
\newline
\indent
Over 250,000 temperature readings were obtained from numerous thermal sensors (thermistors) installed in and around the telescope. The location of sensors relevant to this study are displayed in Figure \ref{fig:schematic}. Three pairs of sensors are placed at three different altitudes in the dome. Half of them are mounted directly on the surface of the metal truss, while the other half are suspended freely in the air next to its corresponding truss sensor. The dome air temperature is calculated by taking the the median of the three dome air sensors. Two sensors are mounted directly on the outer rim of the primary mirror. One is placed on the side that tips down when the telescope slews to lower elevation while the other is placed on the upward side. The average primary mirror temperature is calculated by taking the average of the two mirror sensors. The outside air temperature is recorded by a weather tower located outside of the observatory. Digital readings are taken every 5 minutes and are stored in a database managed by the observatory. 

The relative calibration of the sensors were checked and did reveal offsets. We measured and corrected their offsets 6 months after recalibration by minimizing the residuals between the fit of the calibrated data from the previously uncalibrated points. To ensure offsets were stable we stacked previous years of uncalibrated data. Nevertheless, in this analysis we only study effects due to temperature differences. Therefore absolute offsets should cancel themselves out. Datasets were excluded if there was ongoing GPI or telescope maintenance. 
 
\section{Analysis and Results}
\begin{figure} [!htb]
   \begin{center}
   \begin{tabular}{c} 
   \includegraphics[height=5.5cm]{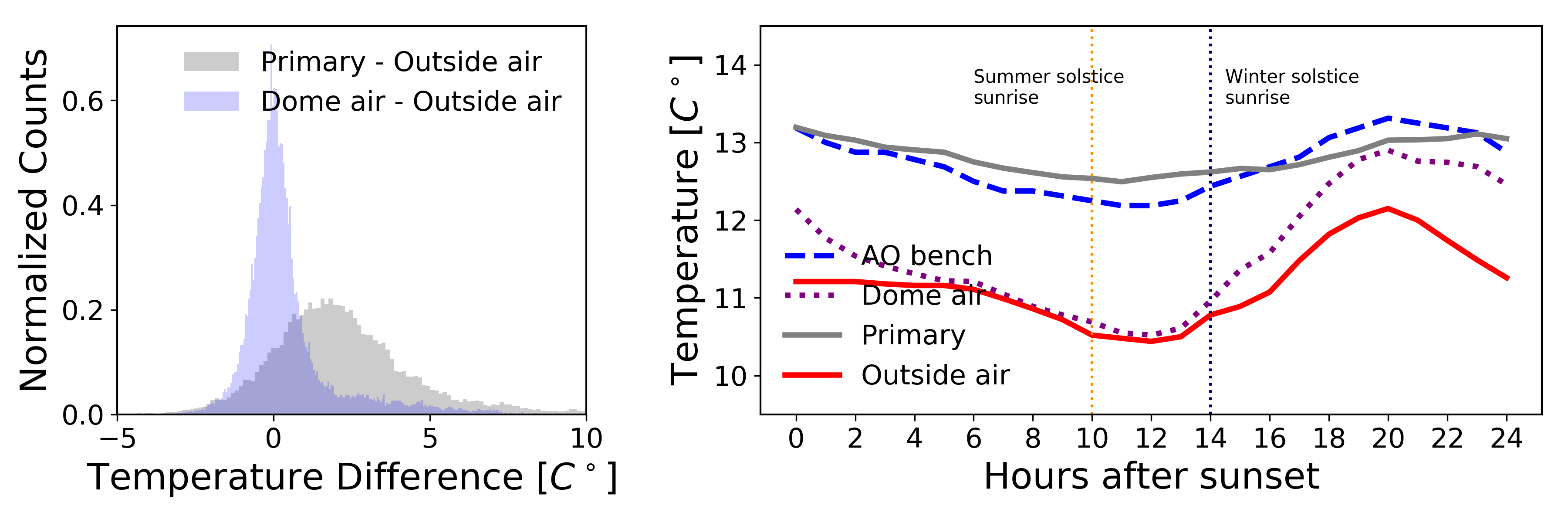}
   \end{tabular}
   \end{center}
   \caption[example] 

   { \label{fig:temphist} 
(Left) Histogram of temperature differences at night. (Right) Temperature vs. number of hours since sunset. Bold lines represent median temperature and dashed lines mark approximate length of the shortest and longest nights. The Primary is constantly 2 C$^\circ$ warmer than the outside air due to its large thermal inertia. GPI's temperature is coupled with that of the primary.}
   \end{figure}

Figure \ref{fig:temphist}(a) demonstrates the challenging thermal conditions at the telescope. In the majority of GPI's observations, the mirror is $\sim 2^\circ C$ warmer than the outside air. This is likely due to daytime heating of the dome air paired with the large thermal inertia of the mirror assembly. Figure \ref{fig:temphist}(b) supports this explanation as it shows that the median primary mirror temperature fails to  converge with the median air temperature. 

\begin{figure} [hh]
   \begin{center}
   \begin{tabular}{c}
   \includegraphics[height=7.5cm]{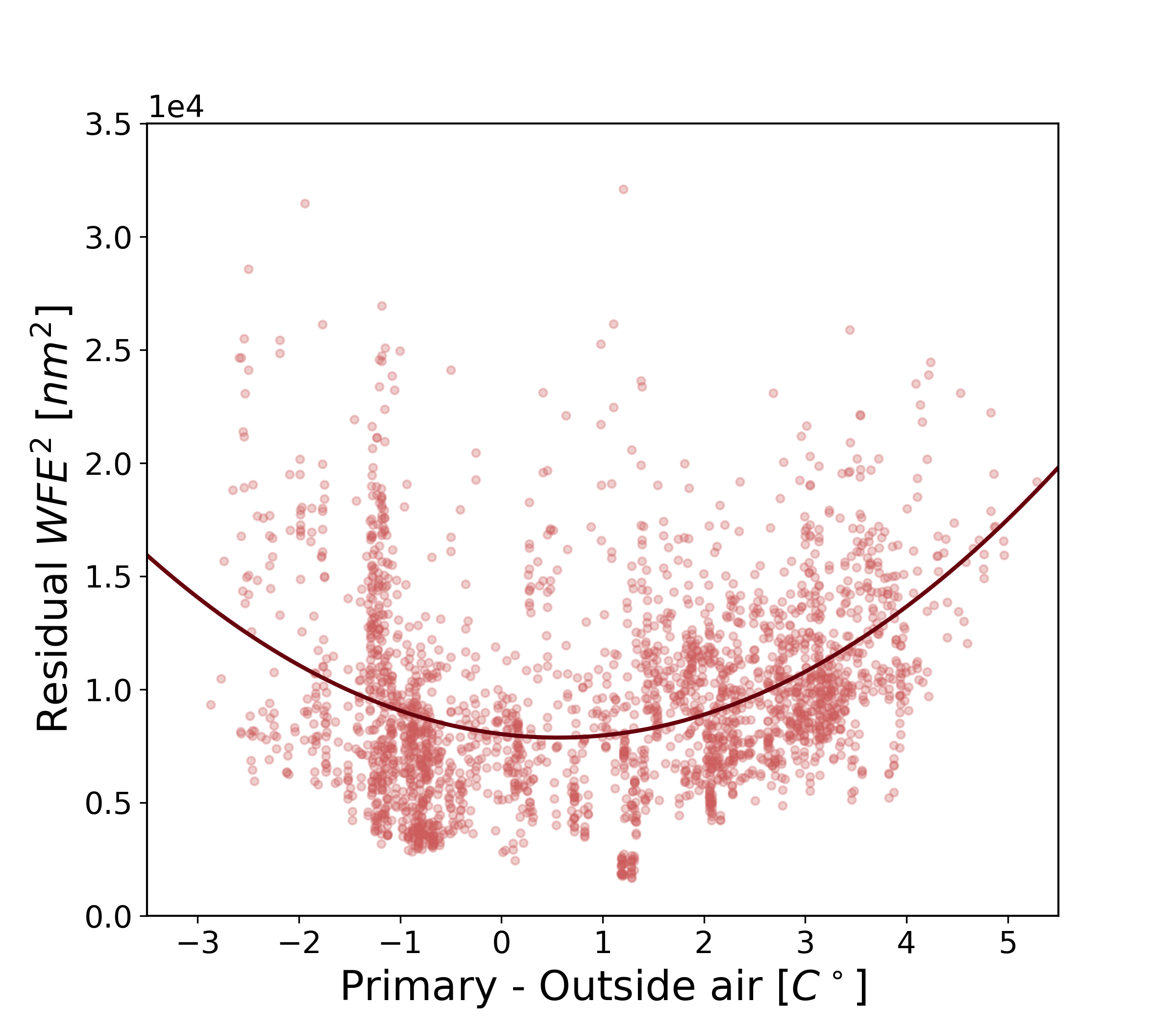}
   \end{tabular}
   \end{center}
   \caption[example] 

   { \label{fig:wfevstemp} 
$\sigma_{WFE}^2$ vs. temperature difference of the primary and outside air during good seeing conditions. Red line indicates the expected trend line as predicted by $\sigma_{WFE}^2 \propto \Delta{T}_M^2$. }
   \end{figure}

The effects of local seeing can be masked by larger contributions from other sources. Thus, we select datasets where error from other effects are reduced. Selecting stars with I-band magnitude $< 7$ ensures that the AO system isn't photon-starved. Selecting datasets observed when $r_o < 1$", and $\tau_o > 1$ ms reduces the error introduced from the atmosphere.  
The residual WFE informs us about the AO system's response to mirror seeing. The Residual $\sigma_{WFE}^2$ versus the absolute temperature difference between the primary and outside air is plotted in Figure \ref{fig:wfevstemp}. Residual WFE grows with mirror-to-air temperature difference. This error can be modeled, with several simplifying assumptions, as the sum of two terms:
\begin{equation}
\label{eq:fov}
\sigma_{WFE}^2 = \alpha \tau_0^{\frac{-5}{3}} + \beta \Delta{T}_m^2 
\end{equation}
Where $\sigma_{WFE}^2$ is the total residual WFE, $\tau_0$ is the coherence time of the atmosphere, and $\Delta{T}_m^2$ is the mirror-to-air temperature difference. We group the data into 17 smaller bins and fit only to the mirror seeing term. The polynomial $y = ax^2 + bx + c$ to the bin averages weighed against the mean error. The model predicts that having a mirror 3 degrees above outside air temperature introduces $\sim$ 52 $nm$ of error. The characteristics of the induced turbulence can be studied in future work by analyzing the slope of the power spectral density function of the wavefront. 

\begin{figure} [ht]
   \begin{center}
   \begin{tabular}{c} 
   \includegraphics[height=7cm]{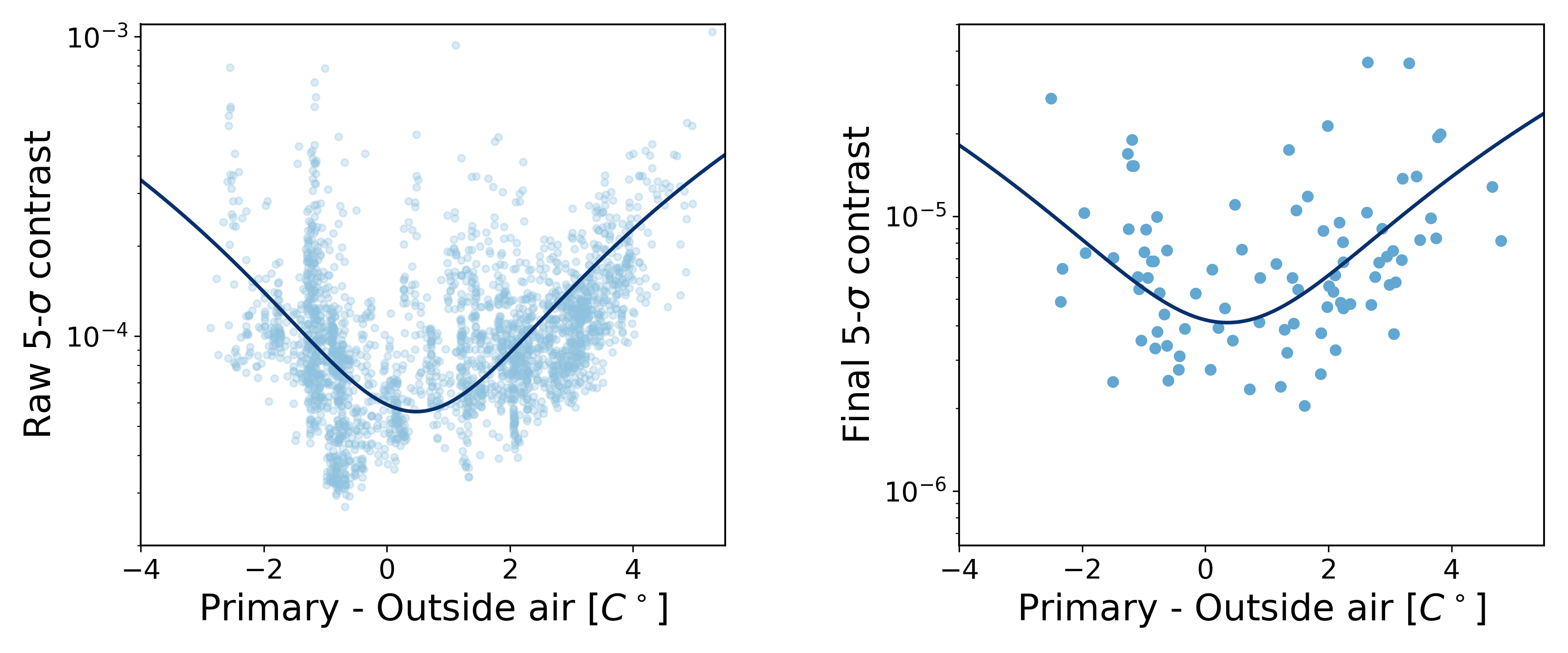}
   \end{tabular}
   \end{center}
   \caption[example] 

   { \label{fig:contrastvsdt} 
Image contrast vs.\ temperature difference between the primary and outside air under good conditions. (Left) is raw contrast while (right) is final contrast. Blue line indicates the model predicted by contrast ∝ Δ 2. On nights when the temperature of the mirror diverged by more than 3 C° from the outside air temperature, contrast degraded by a factor of ~2. }
   \end{figure}

Contrast is the metric we ultimately care about for imaging planets. In Figure \ref{fig:contrastvsdt} we display raw contrast versus the mirror-to-air temperature difference. The quadratic behavior of the data is consistent with the expectation that contrast increases with the square of the Residual WFE. We group the data into 17 half degree bins and fit the polynomial $y = ax^2 + bx + c$ to the average contrast in the bins and weight them against the mean error. Final contrast is more difficult to infer a trend since there a fewer points and it's unclear how post processing will change their values. Nevertheless the fit is consistent with the raw contrast fit.

\section{Conclusion}

We compared the contrast (\textit{i.e.}, detection limit in terms of planet to star flux ratio) of 2,521 60~sec exposures and of 90 post-processed observing sequences from the GPIES campaign  to the simultaneous temperature measurements of the instrument and the Gemini South telescope environment.
We concluded that GPI's sensitivity to planets is directly impacted by temperature differences between the primary mirror and the outside air, which acts as a proxy for atmospheric turbulence inside the dome. Indeed, the performance of the instrument is optimal when the mirror-to-air temperature difference remains small, and the contrast degrades by a factor 2.5 for a temperature difference of $3^\circ$C.
We also showed that the primary mirror constantly stays $\approx 2^\circ$C warmer than the air inside the dome on average, most likely due to its large thermal inertia. We argue that passive cooling from the dome vents is insufficient to ensure thermal equilibrium in the dome. A possible remediation could involve actively cooling the primary mirror during daytime. Also, some studies have shown that slowly blowing a stream of cold air above the mirror surface may also reduce mirror seeing effects\cite{Gillingham1983PROCEEDINGSTelescope}. 

\acknowledgments 
 
The GPI project has been supported by Gemini Observatory, which is operated by AURA, Inc., under a cooperative
agreement with the NSF on behalf of the Gemini partnership: the NSF (USA), the National Research
Council (Canada), CONICYT (Chile), the Australian Research Council (Australia), MCTI (Brazil) and MINCYT
(Argentina). Additionally, portions of this work were performed under the auspices of the U.S. Department of Energy by Lawrence Livermore National Laboratory under Contract DE-AC52-07NA27344. 
V.B. acknowledges government sponsorship; this research was carried out in part at the Jet Propulsion Laboratory, California Institute of Technology, under a contract with the National Aeronautics and Space Administration.

\bibliography{Mendeley_SPIE-_GPI_dome_seeing_}  
\bibliographystyle{spiebib} 

\end{document}